# Learning Enabled Dense Space-division Multiplexing through a Single Multimode Fibre

## Pengfei Fan[1], Michael Ruddlesden[1], Yufei Wang[1], Luming Zhao[1,2], Chao Lu[3], Lei Su[1]*

[1]School of Engineering and Materials Science, Queen Mary University of London, Mile End Road, London, E1 4NS
[2]School of Optical and Electronic Information and Wuhan National Laboratory for Optoelectronics, Huazhong University of Science and Technology, Wuhan 430074, Hubei, P. R. China
[3]Department of Electronic and Information Engineering, The Hong Kong Polytechnic University, Hung Hom, Kowloon, Hong Kong.
*l.su@qmul.ac.uk

**Space-division multiplexing is a promising technology in optical fibre communication to improve the transmission capacity of a single optical fibre. However, the number of channels that can be multiplexed is limited by the crosstalks between channels, and the multiplexing is only applied to few-mode or multi-core fibres. Here, we propose a high-spatial-density channel multiplexing framework employing deep learning for standard multimode fibres (MMF). We present a proof-of-concept experimental system, consisting of a single light source, a single digital-micromirror-device modulator, a single detection camera, and a deep convolutional neural network (CNN) to demonstrate up to 400-channel simultaneous data transmission with accuracy close to 100% over MMFs of different types, diameters and lengths. A novel scalable semi-supervised learning model is proposed to adapt the CNN to the time-varying MMF information channels in real-time, to overcome the environmental changes such as temperature variations and vibrations, and to reconstruct the input data from complex crosstalks among hundreds of channels. This deep-learning based approach is promising to maximize the use of the spatial dimension of MMFs, and to break the present number-of-channel limit in space-division multiplexing for future high-capacity MMF transmission data links.**

The ever-increasing demand for high-capacity data transmission has witnessed multiple innovations in data-channel multiplexing through a single optical fibre in dimensions such as time, wavelength, space, polarization and phase. Among these, space-division multiplexing (SDM), including mode-division multiplexing (MDM), has been exploiting the spatial dimensions of the fibre cross-section for channel multiplexing over the past decade [1-3], and hundreds-terabits-per-second transmission over a hundred-kilometre long fibre was demonstrated recently [4]. The primary technical challenge in SDM is crosstalk management due to the close proximity of the co-existing data channels. Multi-core fibres (MCF), consisting of an array of physically distinguishable single-mode cores in a single fibre cladding, was designed to tackle this problem [5,6]. The individual cores in the MCF are separated with sufficient spacing to serve as independent paths with significantly reduced crosstalks. The total number of channels in MCF is limited, owing to the spacing requirement between cores, and the susceptibility to bending-induced rupture when the cladding diameter is greater than 200μm [7,8]. For MDM transmission in multimode fibres (MMFs), the channel paths overlap significantly, and consequently modes couple randomly during propagation. As a compromise, few-mode fibres (FMF) and multiple-input multiple-output (MIMO) digital signal processing techniques were used for MDM, where only a restricted number of modes, typically 6-12 distinct modes, was utilised for light propagation [9-11]. After that, combining MCF and FMF was proposed to allow higher spatial channel density [3,12,13], and orbital angular momentum (OAM) technique was also demonstrated [14-17].

Deep learning techniques have been successfully applied to fields including computer vision, natural language processing, bioinformatics, drug design, medical image analysis, and material inspection [18-20]. In addition, several groups proposed the application of deep learning in optical communications [21-27]. Recently, it was demonstrated that MNIST digits or other simple geometries can be transmitted through a single MMF via deep learning [28,29]. More generic natural scenes transmission through an MMF up to 10 m was also reported [30]. In particular, we showed that convolutional neural network (CNN) can generalise different MMF transmission states and MNIST digits can be reconstructed at high accuracies even when the MMF is subject to continuous shape variations [29]. To accurately transmit randomly distributed and spatially multiplexed data through a standard MMF over a longer-period of time, such as for optical communication, has not yet been reported, not only because of the difficulty in achieving a high accuracy suitable for communication, but also because of the rapidly varying MMF information channels.

In this paper, we propose a high-spatial-density channel multiplexing framework using deep learning, for randomly-distributed communication-type data transmission over a single MMF. A semi-supervised learning (SSL) model is proposed toward the scalable MMF transmission, to overcome the time-varying nature of the MMF information channels.

**Results:**

The experimental setup used to demonstrate the deep learning based high-spatial-density channel multiplexing framework is shown in Fig. 1a (See Methods section for details) [31]. The transmission experiments are repeated over a range of fibres of different types, diameters and lengths, as detailed in Table 1. For simplicity, the datasets are indexed as follows: 'Refractive index'-'Core diameter'-'Length'-'The number of transmitted channels (denoted by N)'. For example, SI-100-100-400 refers to a step-index (SI) MMF with a diameter of Ø100 μm, a length of 100 m, and 400 multiplexing channels, i.e. $20\times20$ input digital-micromirror-device (DMD) pixels. Fig. 1b is the schematic of the proposed CNN architecture [29], and a standard CNN with three convolutional layers is used as the underlying classifier component (CNN implementation is detailed in Supplementary Section 1).

**Table 1. Experimental Parameters**

| Datasets | Fibres | Description | Data size | CMOS speckles size | N | Frame rate |
|---|---|---|---|---|---|---|
| SI-40-100-25 | SI 40/125-22/250 | Step-index Ø40μm 100m | 40,000 | 320×320 | 5×5 | 200 fps |
| SI-40-100-100 | SI 40/125-22/250 | Step-index Ø40μm 100m | 40,000 | 320×320 | 10×10 | 200 fps |
| SI-100-100-100 | SI 100/140-22/250 | Step-index Ø100μm 100m | 40,000 | 320×320 | 10×10 | 200 fps |
| SI-100-100-400 | SI 100/140-22/250 | Step-index Ø100μm 100m | 40,000 | 320×320 | 20×20 | 200 fps |
| SI-100-1k-100 | SI 100/140-22/250 | Step-index Ø100μm 1km | 40,000 | 320×320 | 10×10 | 200 fps |
| SI-100-1k-225 | SI 100/140-22/250 | Step-index Ø100μm 1km | 40,000 | 320×320 | 15×15 | 200 fps |
| SI-100-1k-400 | SI 100/140-22/250 | Step-index Ø100μm 1km | 40,000 | 320×320 | 20×20 | 200 fps |
| GI-50-100-25 | TechOptics OM3 | Graded-index Ø50μm 100m | 40,000 | 260×260 | 5×5 | 200 fps |
| GI-50-100-100 | TechOptics OM3 | Graded-index Ø50μm 100m | 40,000 | 260×260 | 10×10 | 200 fps |
| GI-50-1k-25 | TechOptics OM3 | Graded-index Ø50μm 1km | 40,000 | 260×260 | 5×5 | 200 fps |

We first test the stability of the MMF information channel. The MMFs under test are bare fibres without protective jackets, winded on a fibre reel as provided by the manufacturers. The stability over time is measured by the correlation to an initial speckle pattern recorded with the same input modulated by the DMD. All MMFs under testing exhibit an unstable time-varying nature due to the environmental changes such as temperature vibrations and vibrations, as shown in Fig. 1c. For example, the correlation drops to be less than 20% through the step-index Ø100μm 1km MMF after 100s. The effects of fluctuations in ambient temperature and airflow are exacerbated by the increase of the fibre length. A periodic fluctuation in the stability of the system is present. Although the significant fluctuations exist in the stability test, Fig. 1d shows the high correlation between two adjacent speckles sampled by the camera along the time domain. This is an important observation, revealing that the system only changes gradually, which forms the basis of our SSL model to be introduced later.

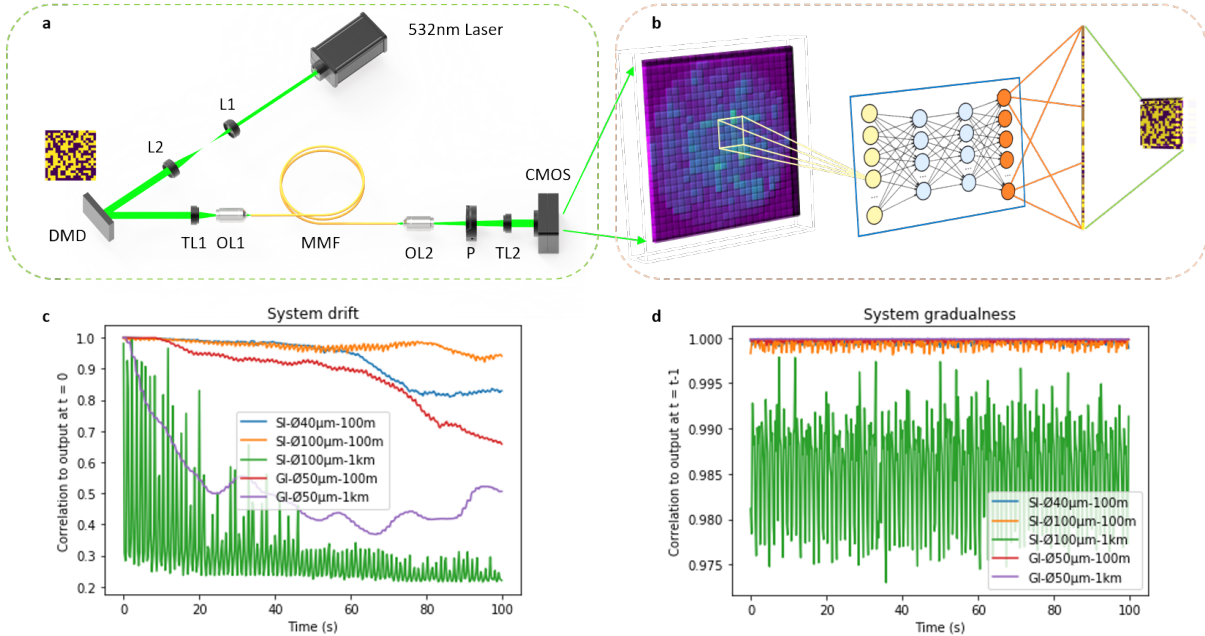

**Fig. 1. The proposed high-spatial-density single-MMF channel multiplexing framework and the time-varying nature of the MMFs under test. a.** The proof-of-concept experimental system. The beam is expanded, collimated and directed onto the DMD, the reflection of which is coupled into the MMF. The output from the MMF is captured

by a CMOS camera. L1, L2: Bi-Convex Lenses; TL1, TL2: Tube Lens; DMD: Digital Micromirror Device; OL1, OL2: Objective Lens; MMF: Multimode Fibre; P: Polarizer; CMOS: Camera. **b**. The proposed CNN architecture to reconstruct the transmitted data. For simplification, some layers are not shown. **c**. Stability of the optical system over time for different fibres, represented by the correlations between the output speckle at different time intervals to the initial output of the same input. **d**. The strong temporal dependency of adjacent speckles, represented by the correlation between the two adjacent output speckles of the same input.

To verify the idea, we conduct the experiment in an offline mode first, where all training data are collected altogether, and the evolving transmission states (i.e. instability) of the MMF are learned jointly by the CNN at the same time in a single training process. For each dataset described in Table 1, 90% (36,000) of these speckles and their corresponding DMD patterns are randomly selected for training with the remaining 4000 data pairs for testing the final CNNs. In this offline setting, the CNNs can generalise the instability very well in all of these drifting transmission states. Examples of the data transmission accuracies of our proposed framework are shown in Fig. 2a-2d for Datasets: SI-40-100-100, SI-100-100-400, SI-100-1k-400 and GI-50-100-100. Examples of the prediction results are shown in Fig. 2e. The complete results for all other datasets are provided in Fig. S4 in Supplementary Information Section 3. Using CNNs, we are able to effectively multiplex the inputs of up to 400 data channels by detecting the corresponding speckle patterns, with high prediction accuracies for the datasets listed in Table 1. Despite the high instability with the 1 km fibres, the prediction accuracy remains good (e.g. over 99% for the multiplexing of 225 channels). This suggests that the CNN generalises the instability in its learning and is hence able to achieve prediction accuracies similar to those achieved with short fibres. As shown in Table 2, for the multiplexing of 100 channels, an increase in step-index Ø100μm fibre length from 100m to 1km yields only a ~1.2% drop in prediction accuracy, and doubling the step-index 100m fibre diameter results in a further 100% prediction accuracy. A drop in prediction accuracy of most datasets is clearly visible towards the start and the end of the collection. This is due to the reduced number of input-output data pairs in the training datasets at the start and the end. This is because, at the starting point, there are no immediate correlated training data available before this point, and similarly at the endpoint no immediately correlated training data available after the endpoint either. Therefore, the relatively fewer training datasets for the MMF transmission states at the start and the end of the data collection in our laboratory testing result in less accurate predictions, which are not the real system performance and can be disregarded.

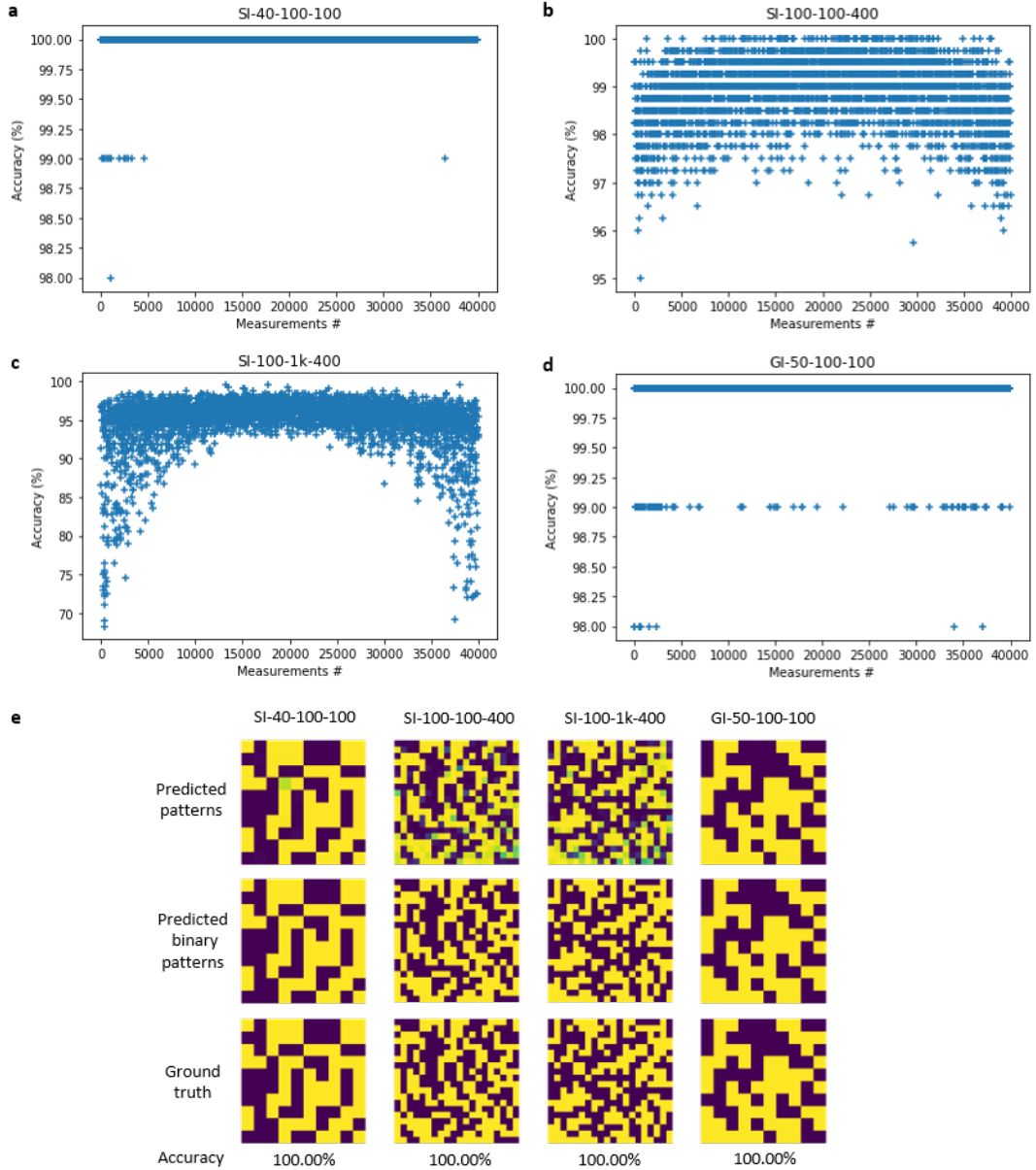

**Fig. 2. Experimental results in an off-line setting using Datasets SI-40-100-100, SI-100-100-400, SI-100-1k-400 and GI-50-100-100 as examples. a-d**. Experimental prediction accuracies for 10% (4000) randomly selected test pairs for different datasets. These show the prediction accuracies at their location with regards to all 40,000 input-output pairs obtained. **e**. Examples of predicted patterns and their corresponding accuracies for different fibre lengths and acquisition methods. The drops in prediction accuracy at the beginning and the end of the measurement in **a-d** are caused by the reduced number of training data at the beginning and the end (detail discussions are in the text).

**Table 2. Average accuracy, F1_score, MSE, PCC, SSIM (defined in Methods section) and output runtime for different datasets**

| Datasets | Accuracy (%) | F1_score | MSE | PCC | SSIM | Network output runtime (*sec*) |
|---|---|---|---|---|---|---|
| SI-40-100-25 | 99.9990 | 1.0000 | 0.0001 | 1.0000 | 1.0000 | 0.0002 |
| SI-40-100-100 | 99.9958 | 1.0000 | 0.0001 | 0.9999 | 0.9999 | 0.0002 |
| SI-100-100-100 | 100.0000 | 1.0000 | 0.0001 | 1.0000 | 1.0000 | 0.0002 |
| SI-100-100-400 | 98.8528 | 0.9884 | 0.0090 | 0.9819 | 0.9872 | 0.0002 |
| SI-100-1k-100 | 98.7578 | 0.9872 | 0.0101 | 0.9793 | 0.9759 | 0.0002 |
| SI-100-1k-225 | 99.0281 | 0.9902 | 0.0078 | 0.9841 | 0.9816 | 0.0002 |
| SI-100-1k-400 | 94.9149 | 0.9489 | 0.0385 | 0.9193 | 0.9322 | 0.0002 |
| GI-50-100-25 | 99.9930 | 0.9999 | 0.0001 | 0.9999 | 0.9998 | 0.0002 |
| GI-50-100-100 | 99.9700 | 0.9997 | 0.0003 | 0.9994 | 0.9998 | 0.0002 |
| GI-50-1k-25 | 99.9950 | 0.9999 | 0.0001 | 0.9999 | 0.9998 | 0.0002 |

The offline experiment detailed above proved that high transmission accuracy can be achieved using one CNN to generalise the instability in the MMF. However, the offline setting is not suitable for real-world optical fibre communication. Firstly, it is impossible to acquire all the training data altogether for the time-varying MMF information channel. Secondly, there is no direct access from the receiver to the data at the transmitter, and therefore the CNN training data cannot be directly gained during the real-time optical communication. To address this without sending known channel calibration data, we note that the offline results presented in Table 2 suggesting ~100% accuracy for all the datasets, and the MMF information channel only varies gradually with time (Fig. 1d). Therefore, it is possible to update the CNN in real-time by using the CNN predicted channel inputs and camera-collected corresponding speckles.

We propose a confidence-based SSL model (see Methods section for details) to overcome the significantly time-varying nature of MMF transmission channels. In this SSL model, we only use a small batch of initial labelled data according to extreme verification latency (EVL) [32,33] conditions in an online setting, and then we include the real-time transmitted data as pseudo labels in the training process to overcome the predictive errors caused by the time-varying MMF channels. This is a receiver-end framework which requires no feedback from the transmitter after the initialisation. The experiments are conducted using six datasets collected with different fibres: SI-40-100-25, SI-40-100-100, SI-100-100-100, GI-50-100-25, GI-50-100-100, and GI-50-1k-25. Each of the datasets consists of 40,000 input and output pairs. In Fig. 3, the performance of our confidence-based SSL model is compared with the performance of a static model, where the CNN is only trained with the initial labelled instances without being further updated over time.

In Fig. 3a, with 25 channels transmitted in a 40μm-core 100m step-index MMF, significant fluctuations can be seen from the results of the static model, compared to a ~100% accuracy achieved by our SSL model. Fig. 3b and 3c are results arising from the multiplexing of 100 channels over 40μm and 100μm core 100m step-index MMFs respectively, where our SSL model achieves 100% accuracy consistently over the experiment duration. In comparison, without using our SSL model, the accuracy starts to decrease before Step 10 and drops quickly (the Step describes the process of using of a new batch of input-output pairs to update the CNN. See Methods Section for the definition of Step). This implies that the degradation of performance is caused by accumulated predictive errors. Fig. 3d and 3e illustrate that the proposed SSL algorithm achieves near 100% accuracy over time for the multiplexing of both 25 and 100 channels over a 50μm-core 100m graded-index MMF. Without using the SSL model, the transmission accuracy decreases drastically over time. Fig. 3f displays the SSL model results of multiplexing 25 channels over a graded-index MMF of 1km length. Again, our SSL model exhibits 100% transmission accuracy over time.

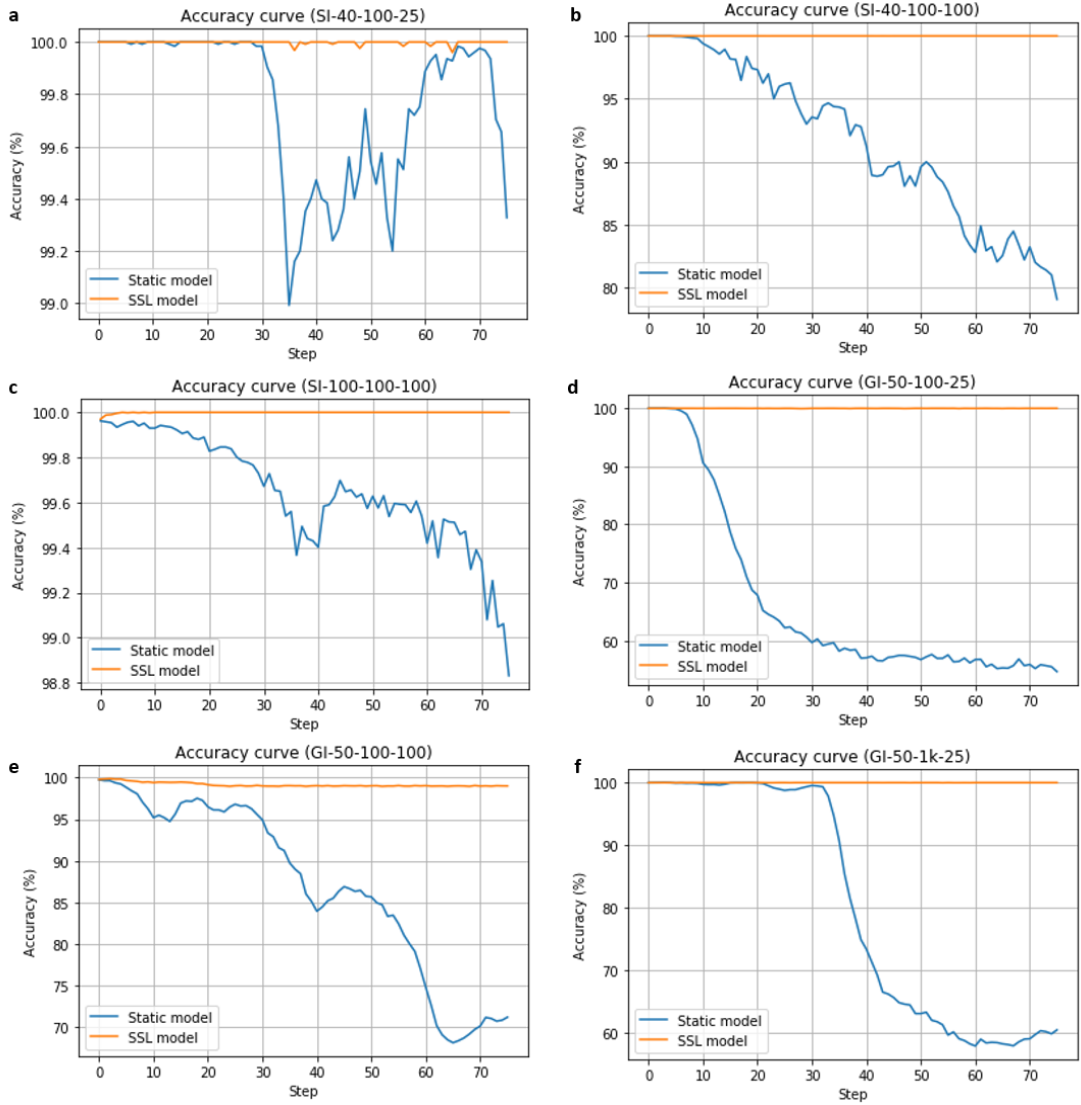

**Fig. 3 Performance of our confidence-based SSL model for Datasets**: **a**. SI-40-100-25, **b**. SI-40-100-100, **c**. SI-100-100-100, **d**. GI-50-100-25, **e**. GI-50-100-100, and **f**. GI-50-1k-25. Each 'Step' is the use of a new batch of input-output pairs to update the CNN during the transmission, as detailed in the Methods Section.

To evaluate the performance of individual data channels for results presented in Table 2, we select four datasets with relatively low data transmission accuracy, namely SI-100-100-400, SI-100-1k-100, SI-100-1k-225 and SI-100-1k-400. Over 4000 predicted inputs, the transmission accuracy of each individual channel is calculated by the total number of accurately transmitted channel data divided by 4000. Fig. 4 visualises individual channel transmission accuracy for these four datasets. It is interesting to see from Fig. 4, that the individual channel performance is not evenly distributed. Certain channels perform much worse than others. Surprisingly, these less accurate transmission channels remain at similar spatial locations of the DMD modulated pattern for experiments on different MMFs. For example, Fig. 4a and d are the channel performance results for the multiplexing of 400 channels over 100m and 1km SI MMFs, respectively, and the Pearson correlation coefficient between them is as high as 0.898. Note that each individual channel occupies the same number of DMD pixels at the input regardless of the number of channels multiplexed (see details in Methods section). Therefore, we consider the zoom-in areas in the red-lined squares in Fig. 4c and d cover the similar DMD area of Fig. 4b. The calculated Pearson correlation coefficients between Fig. 4b and the red-lined square areas are reasonably high too, 0.759 and 0.763 respectively. This suggests that different experiments suffer from identical system noises, which are likely caused by the experimental system rather than the MMF. Hence, the transmission accuracy can be further enhanced by improving the experimental system or by excluding those less-accurate channels.

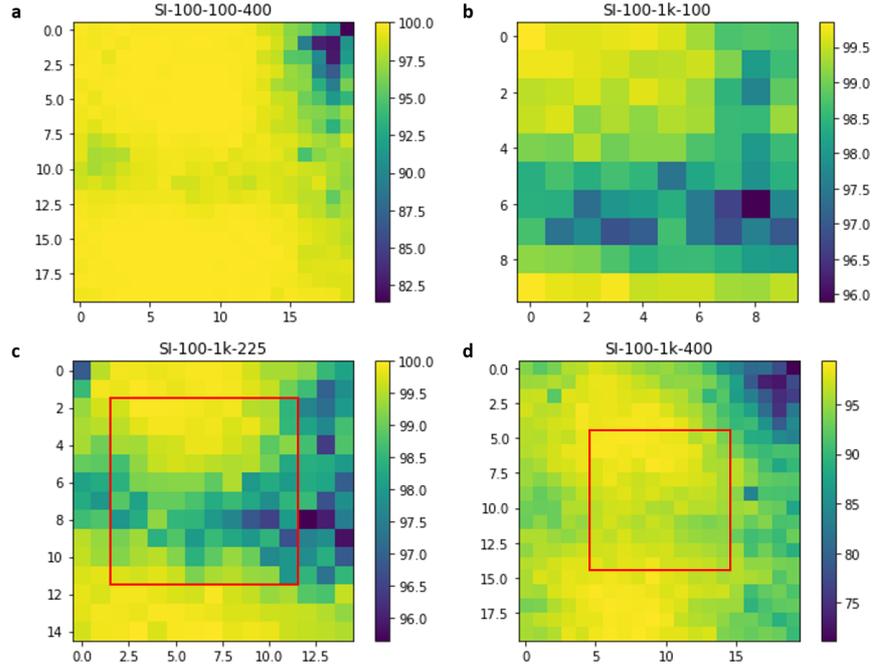

**Fig. 4 The individual transmission channel performance for Datasets: a**. SI-100-100-400, **b**. SI-100-1k-100, **c**. SI-100-1k-225 and **d**. SI-100-1k-400. The transmission accuracy of individual channel is defined as the percentage of accurately transmitted data of each individual channel.

Finally, we study the influence of the total number of pixels at the receiver camera end on the transmission accuracy, as shown in Fig. 5. Firstly we down-sample the original speckles to speckles with reduced numbers of pixels (see Methods section), as illustrated in Fig. 5a. It can be seen in Fig. 5b, that a down-sampled speckle at 30×30 pixels is sufficient to recover the full-transmission-channel information accurately for SI-40-100-100 and SI-100-100-100. With the increase of the multiplexed channel numbers and the MMF length, it appears more camera pixels are required to maintain a high detection accuracy. For example, approximately 70×70 and 80×80 detector pixels are needed for accurate data transmission of Datasets SI-100-100-400 and SI-100-1k-100, respectively. For SI-40-100-25, only 10×10 detector pixels are needed to achieve a transmission accuracy of more than 98%. These results suggest that the total pixel number can be optimised at the receiver end to further enhance the detection and transmission speed in real communication data links. We also studied other partial speckles and their influences on the accuracy, and presented the results in Fig. S6 to Fig. S8 in Supplementary Section 3.

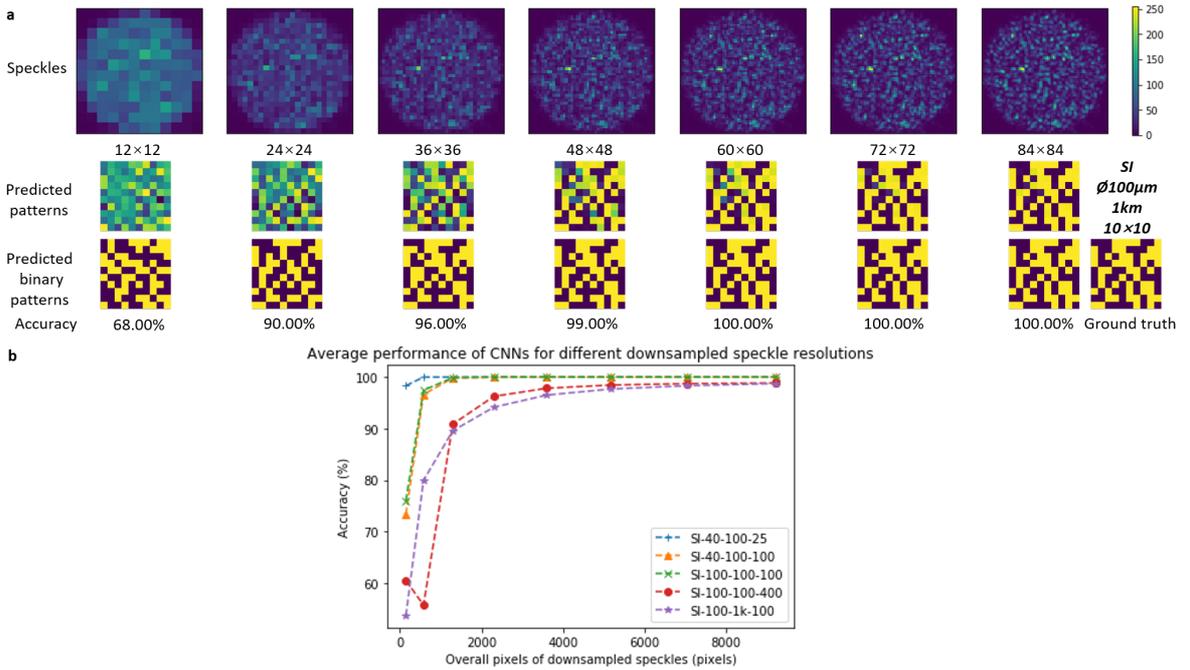

## Discussion:

There are a few limiting factors in our proof-of-concept experiment. Firstly, the modulation speed of DMD (~22kHz) and the frame rate of the camera (200f/s) used in this work are much less than those of the optical fibre modulators and the photodetectors used in optical communication. In fact, the transmission speed of our experiment is limited by the camera frame rate. Given the speed limit of our proof-of-concept system, the intermodal dispersion with the MMF is not studied and the input-output data are collected through a trigger signal. This could be addressed by using an ultra-high-speed camera or a photodetector array. As shown in our results, a down-sampled speckle is sufficient to recover the complete channel information. At the transmitter end, multiple modulators and light sources can be used to replace the DMD for a much faster speed. Secondly, the error rate is still higher than the bit error rate expected in the real optical fibre communication. We envisage that by involving the error correction techniques widely used in optical communication, such as forward error correction (FEC) technologies [34], the transmission accuracy can be improved within the acceptable range. In addition, as suggested by the results presented in Fig. 4, channel performances are affected by the experimental system. Hence, transmission accuracy can be further improved with an optimised system or by excluding those less-accurate data channels. We are also aware of the fact that the online training and updating of the CNN in real-time may be time-consuming and challenging for ultrafast data communication, and a practical training and updating scheme needs to be carefully designed and developed for real communication applications. With the rapid development of artificial intelligence and high-performance computing technologies [35,36], a negligible time frame of neural network training and prediction may be within the reach in the future.

In summary, we presented a high-spatial-density multiplexing transmission framework over MMFs using deep learning. Our proof-of-concept experiments demonstrate that deep learning using CNN enables high-spatial-density channel multiplexing for accurate data transmission of up to 400 channels over a single MMF. The highly time-varying MMF transmission is overcome by an SSL algorithm, which updates the CNN in real-time according to the changes of the MMF information channels to maintain an accurate transmission over time. The proposed high-spatial-density channel multiplexing framework has the potential to become a promising technology for future high-capacity data links over multimode optical fibres.

## Methods:

### Experimental setup:

The laser beam (532nm, 50mW, Cobolt Samba) is expanded, collimated, and projected onto a DMD (ViALUX V-7001, >20kHz). The DMD is able to display any arbitrary binary pattern by switching the micromirrors 'ON' or 'OFF', corresponding to input binary values '1' or '0' respectively. The wavefront of the incident laser beam is modulated by the pattern on the DMD and is consequently coupled into the input of a single solid-core MMF using a tube lens and microscope objective. Different types of MMFs are used in our experiment as listed in Table 1. At the distal end of the fibre, a further microscope objective and tube lens are used prior to image the speckle pattern on a CMOS camera (QImaging optiMOS). In the experiments, the effective area of the DMD (the area coupled into the MMF as the input) is varied accordingly to the total number of transmission channels. Each transmission channel occupies 4×4 DMD micromirror pixels. For example, 20×20, 15×15, 10×10 and 5×5 transmission channels can be obtained by maintaining the total micromirror pixel numbers at 80×80, 60×60, 40×40, 20×20 respectively. The data frames displayed on the DMD are randomly generated (implemented using *random.randint* function within the Python Numpy Library) in an N×N square configuration [37], with 50% each 'ON' and 'OFF' pixels, which resembles the data distribution during real data transmission over N×N independent channels. MMF output speckles are recorded on the CMOS camera over a cropped region of interest measuring 320×320 pixels for step-index MMFs and 260×260 pixels for graded-index MMFs separately.

### Data preprocessing and evaluation metrics:

The system operates at 200 input-output-pairs per second, and 40,000 data pairs are collected during the 200s period. The images are then down-sampled to 96×96 pixels and the dynamic range is subsequently normalised to the range of [0, 255], in order to reduce the number of deep CNN parameters and computer memory usage, therefore increasing the training speed with little degradation to the prediction results. The down-sampling process is implemented using *resize* function with *INTER_AREA* interpolation within the Python cv2 Library.

In offline testing (Fig. 2), the 4000 test input-output data pairs are pseudorandomly selected using *train_test_split* function within the Python Scikit-learn Library by setting the random state to a fixed value, which enables us to relocate the test data as their original orders in the 40,000 measurements. The performance of the models can be numerically quantified from two perspectives [29]: classification evaluation and regression evaluation. Regarding classification evaluation, the accuracy and the $F_1$_score are calculated between the predicted binary pattern and the ground truth. The prediction accuracy is defined as the percentage of correctly predicted pixels within one $N \times N$-pixel input DMD pattern. On the other hand, for regression evaluation, the mean squared error (MSE), the Pearson correlation coefficient (PCC) and the structural similarity index (SSIM) [38] are also applied between the predicted pattern and the ground truth.

**Confidence-based SSL algorithm:**

As illustrated in Fig. S1 in Supplementary Section 2, the proposed confidence-based SSL framework consists of six processing steps: initialising, training, predicting, weighting, filtering and abstaining. The framework begins from the initialising phase, in which the labelled data are provided only once due to the EVL assumption. The second phase is to train a CNN model using the current labelled or pseudo labelled data. And then, the trained model is used to predict the labels of the incoming unlabelled batches in the third step. After predicting phase, the fourth step is to weight the predicted instances using an algorithm monitoring the confidence level of each incoming instance. The fifth step is to filter the most confident samples by cutting the weighted instances using an excluding percentage which is calculated dynamically. The last step is the abstaining phase, in which the classifying process applied to the selected instances is allowed to abstain from making a final decision according to a self-adaptive threshold. After all these steps, the procedure returns the instances with pseudo labels to the second phase for the next round training. It should be noted that a sliding window for training is maintained in our framework to filter the predicted incoming samples and retain the old instances according to a dynamic excluding percentage. In the experiments, the window length is set as the size of the initial labelled dataset. We also release the source code of our confidence-based SSL model for reproducibility purposes [39]. More details of the proposed confidence-based SSL framework, parameter optimization, and training procedures are provided in Supplementary Information.

For calculating the results presented in Fig. 3, the streaming data are received in a chronological order [40-43] starting with only 5% (2000) initial labelled instances, with the remaining unlabelled data stream divided into 76 batches of 500 output speckles in each step. The length of the sliding window is set equal to the number of initial labelled instances, that is, maintaining a buffer with the 2000 most recent samples. Due to the evolving nature of the streaming data, it is necessary to apply prequential metrics [44,45] that are measured sequentially in chronological order. As shown in Fig. 3, the performance of the models quantified by the prediction accuracy is computed. Other average prequential metrics are also applied and detailed in Supplementary Section.

**Data availability**

All data generated and analysed during this study are available from the corresponding author on reasonable request.

**Contributions**

L.S. conceived the idea and designed the experiment. M.R. set up and performed the experiments with the contribution from Y.W.. P.F. designed the neural network and performed data analysis. L.S. and P.F. analysed the results. L.S., P.F. and M.R. wrote the manuscript with the contribution from all authors. L.S. guided the work. C.L. provided guidance and advice on optical fibre communication. L.Z. provided advice and support on multimode optical fibres.

**Acknowledgement**

This work was supported by Engineering and Physical Sciences Research Council (grant number EP/L022559/1 and EP/L022559/2); Royal Society (grant numbers RG130230 and IE161214); H2020 Marie Skłodowska-Curie Actions (790666). This research utilised Queen Mary's Apocrita HPC facility, supported by QMUL Research-IT. http://doi.org/10.5281/zenodo.438045

**Competing interests**

The authors declare no competing interests.

## Supplementary Information:
## Learning Enabled Dense Space-division Multiplexing through a Single Multimode Fibre

### 1. Convolutional neural network (CNN) implementation

Fig. 1b in the main text describes the simplified schematic of the proposed CNN architecture [1]. We consider this binary pattern prediction task a multi-binary-classification problem, and a standard CNN with three convolutional layers is used as the classifier component to prove our concept. The overall structure consists of three convolutional architectures and a fully-connected architecture. Each layer in the convolutional architecture gains feature maps of its previous layer and feeds its own feature maps to the subsequent layer. In our implementation, the convolutional block compromises four sequential operations: a 3×3 convolution, activation (ReLU), batch normalization (BN) [2] and an optional dropout [3]. The numbers of feature map used for the three convolutional layers is 32, 64 and 64 respectively. The strides of convolution performed empirically in the network are 1, 2 and 1 severally. The final convolutional block is followed by a flattening layer. This flattening layer yields one-dimensional flatten units, which are subsequently activated by a ReLU activation layer. After the last BN and dropout layers, we implement a dense layer, which forces the architecture to maintain the exact output dimensions as the labels. The activation function used in the dense layer is a sigmoid function which bounds the output to the 0-1 range. Since the outputs of the CNN are continuous values, a threshold of 0.5 is used to make a final binary decision.

Due to the applied sigmoid output layer for the multi-binary-classification purpose, we here use Binary Cross Entropy cost as the loss function, which can be expressed as:

$$L = -\frac{1}{n_{bs}} \sum_{i=1}^{n_{bs}} \sum_{j=1}^{N} (t_{i,j} \ log \ (sigmoid(\boldsymbol{I}_i, j)) + (1-t_{i,j}) \ log \ (1-sigmoid(\boldsymbol{I}_i, j))) \ , \tag{S1}$$

where $t_{i,j}$ and $(\boldsymbol{I}_i, j)$ define the expected target (label) and the logit of the predicted output on the $j$-th pixel value of the DMD pattern corresponding to the training image $\boldsymbol{I}_i$, i.e. $\boldsymbol{t}_i = [t_{i,1}, \cdots, t_{i,N}]$. $N$ is the macro pixels of DMD pattern, and $n_{bs}$ denotes the batch size. In our implementation, targets and logits must have the same type and shape. Following the CNN's loss function minimization, the error between the output pattern and its corresponding ground-truth is back-propagated through the network and a stochastic *AdaDelta* optimization [4] is used to optimize the CNN's parameters. In our implementation, the learning rate parameter is initially set as 0.01, and the total batch is split into mini-batches of 32 patches each. The entries of 3×3 element kernels used in convolution are initialized using *glorot_uniform* [5]. All bias terms are initialized as 0. The dropout rate is determined as 0.4 for good performance. The validation loss is used as a monitor for the learning rate reducing mechanism and the early stopping strategy where the two patience arguments are set to 7 and 10 separately during training. If the validation loss does not decrease in the last 7 epochs, the initial learning rate is lowered by a factor of 10 until a set minimum value (here we use $1 \times 10^{-7}$) is reached. If the number of epochs with no improvement reaches 10, the training will be stopped. The min-delta argument of $1 \times 10^{-6}$ is set to the minimum change in the monitored quantity to qualify as an improvement. Note that the learning rate reducing mechanism and the early stopping strategy are only used in off-line setting and that the learning rate and the number of training epochs are fixed in the on-line scenarios. The CNN architecture is implemented using the Keras library and the deep neural network is set up by using TensorFlow back-end (Google). We utilise Queen Mary's Apocrita HPC facility to train our network's models. The training phase of the network was performed under the Linux Singularity container using a Tesla V100 GPU card (Nvidia) and a 16 Core Xeon Gold 6142 processor (Intel) with 7.5GB RAM requested for each core.

### 2. Confidence-based semi-supervised learning (SSL) framework

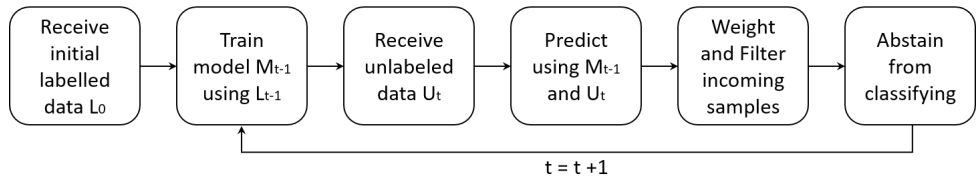

Fig. S1. The proposed confidence-based SSL framework.

The proposed framework is derived from data cleaning concept, i. e. cleaning the error-prone data and feeding a reliable dataset that is free of errors to the training algorithm. The primary aim of data cleaning is to identify errors. To clean the data for training, the confidence level of each upcoming instance is being monitored, and only the most confident samples remain into the training dataset. As an example demonstrated in Fig. S2, the output of each pixel in the predicted pattern is a continuous value from 0 to 1, which denotes the posterior probability of the binary classification. Since the decision threshold is determined as 0.5, this probability value reveals its confidence in the decision made by the model as the final classification 0 or 1. Here we directly leverage the posterior probability as its confidence level to indicate the intrinsic predicting quality of each pixel regardless of its original label. The average confidence level of one predicted instance is calculated as:

$$confidence = \frac{1}{n} \sum_{i=1}^{n} \left( |p_i - 0.5| + 0.5 \right), \qquad (S2)$$

where $p_i$ defines the posterior probability (continuous value) of each pixel from the predicted pattern and $n$ is the number of channels of the DMD pattern.

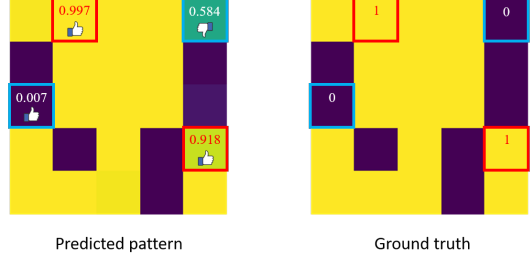

Fig. S2. An example of the predicted pattern to show the confidence level of predicting quality.

The weighted instances in the incoming batch are ranked by their corresponding confidence levels. Along with a dynamic excluding percentage $\alpha$, only the most confident samples in the current batch are selected. The number of samples to be deleted in the on-line process is determined by the excluding percentage $\alpha$, which depends on the degrees of system drifts. The calculation is under the assumption below: if the system drifts significantly, more predicted instances with pseudo labels should be discarded to eliminate the effect of the predictive error as the current model fails to generalise the incoming data batch accurately.

In our method, the system drifts are statistically measured by the squared Hellinger distance between the previous and the current distributions of each attribute in the speckles from two subsequent batches. The varying system results in that the statistical distribution of output speckles suffers changes over time. The squared Hellinger distance is used to quantify the similarity between two probability distributions [6]. For two discrete probability distributions $P = (p_1, \cdots, p_k)$ and $Q = (q_1, \cdots, q_k)$, their squared Hellinger distance is defined as:

$$H^2(P,Q) = \frac{1}{2} \sum_{i=1}^{k} \left( \sqrt{p_i} - \sqrt{q_i} \right)^2. \qquad (S3)$$

As illustrated in Fig. S3, $Batch1$ and $Batch2$ are two batches of speckles from random DMD patterns received in two subsequent time intervals. For the $(i,j)$-th attribute (pixel) in the speckles, the squared Hellinger distance $h_{ij}^2$ is calculated between the distributions $P$ and $Q$ of the intensity value. It iterates through $n \times n$ number of pixels in the speckles, creating two intensity vectors $u$ and $v$ with the $(i,j)$-th pixel from two batches. The two distributions $P$ and $Q$ are characterised using frequency derived from the histograms $h_u$ and $h_v$ computed from the intensity vectors $u$ and $v$, separately. The square root of the number of instances in one batch is set as the number of bins in the histograms. The two histograms are aligned for a reasonable comparison, by assigning the first edge and the last edge of the bins as the minimum value and the maximum value from the two intensity vectors $u$ and $v$. After that, the average distance $H$ through $n \times n$ number of pixels between two batches is calculated and set as the excluding percentage $\alpha$. Finally, the algorithm constrains the excluding percentage $\alpha$ within a determined lower and upper bound pair, $\alpha_1$ and $\alpha_2$ separately.

The squared Hellinger distance is an instance of $f$-divergence. Unlike the $KL$-divergence, it yields a bounded symmetric metric on the space of probability distributions. The factor $1/2$ in front of the calculation forces the squared Hellinger distance to be normalised and range from 0 to 1, in which case it directly represents a percentage coefficient to the drift detection.

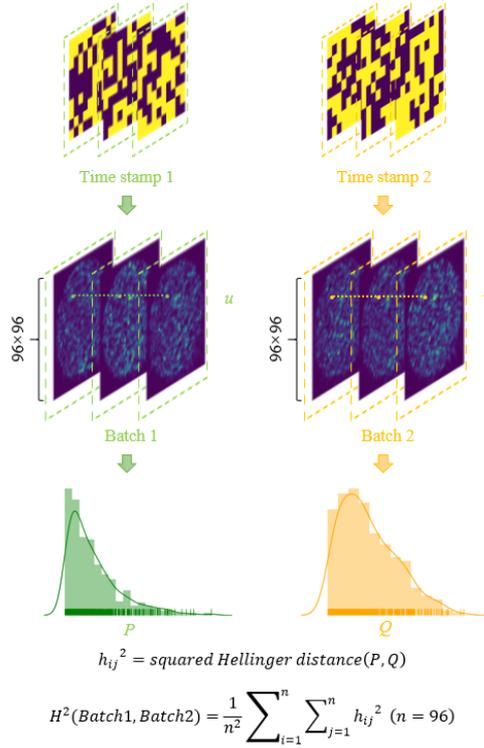

$$h_{ij}^2 = squared\ Hellinger\ distance(P,Q)$$

$$H^2(Batch1, Batch2) = \frac{1}{n^2}\sum_{i=1}^{n}\sum_{j=1}^{n} h_{ij}^2 \quad (n = 96)$$

Fig. S3. Schematics of the squared Hellinger distance calculation.

In Fig. S2, the predictive error occurs when the model classifies the value in the upper right corner in the pattern using a fixed threshold of 0.5. The threshold determines 0.584 as 1, making its final decision arbitrarily regardless of its ground truth. The last step of the learning procedure allows the classifying process to abstain from making its final classifying decision if the certainty displayed by the model is below a specified threshold. Using a fixed threshold degrades the prediction performance because of the drifting data stream. We introduce a self-adaptive threshold which changes dynamically according to the increase or decrease of mean confidence level compared between the previous and current batches. The dynamic abstaining process is firstly initialized with a threshold $\theta_0$. If the mean confidence level of the current batch decreases compared with the previous confidence value, then the abstaining threshold will increase by an adjustment factor $\Delta\theta$ to prevent the model from making decision arbitrarily and blindly, and vice versa. Meanwhile, the threshold $\theta$ is monitored to make sure it is between the range that we stipulate, thus it should be greater than $\theta_1$ and lower than $\theta_2$.

Hence for each channel of the predicted pattern, we will only classify those pixels with certainty satisfying the current threshold restrictions. As the confidence level of each instance has been calculated in the dynamic excluding process, the proposed abstaining strategy doesn't impose any heavy computational costs in the overall procedure. It is worth to notice that this dynamic abstaining threshold is applied only when the instances with pseudo labels are included into the training dataset and that for the primary data reconstruction task to the receiver end, the threshold of 0.5 is still used to generate the final outputs.

## 3. Extended results

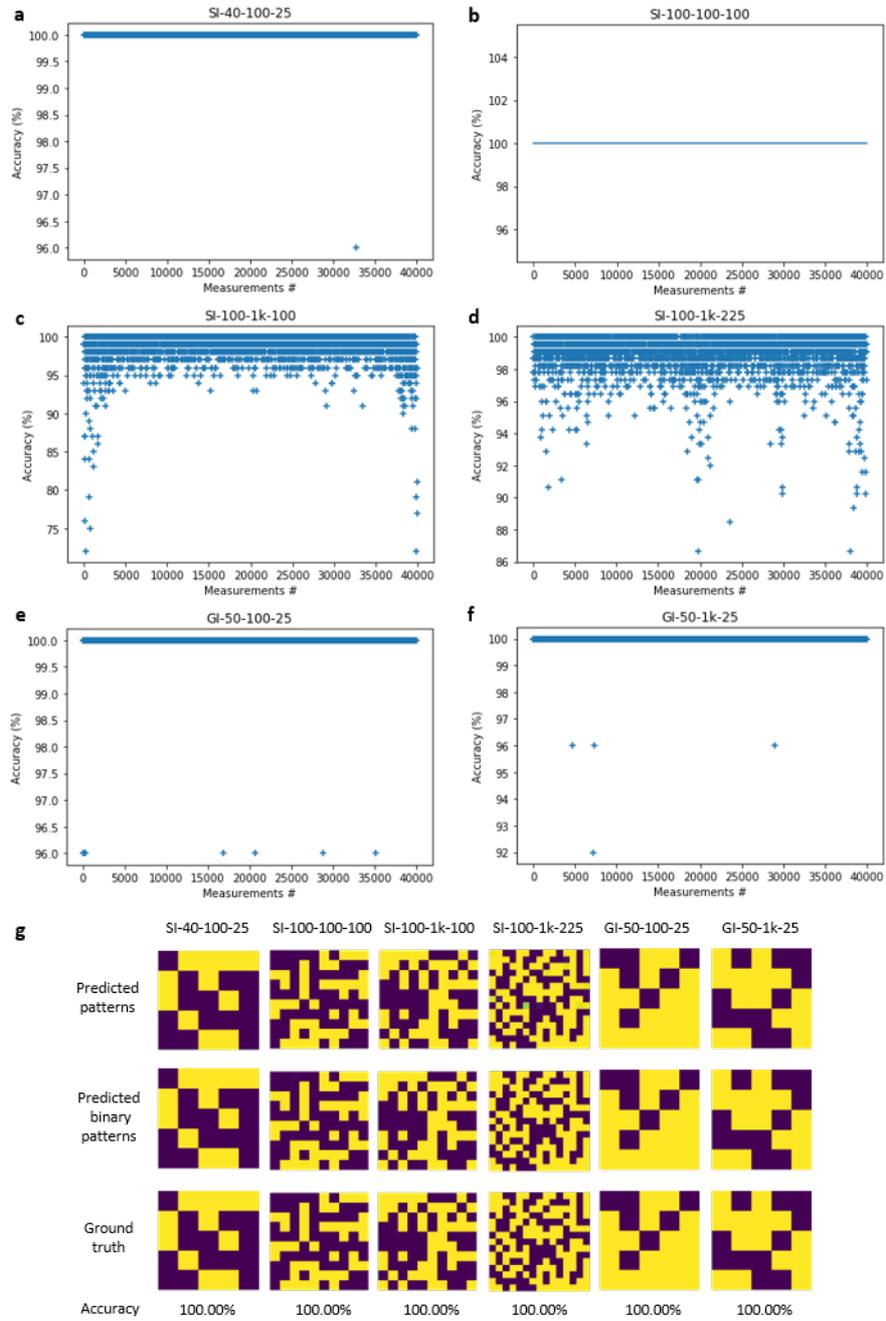

Fig. S4. Experimental results tested in an offline setting using Datasets SI-40-100-25, SI-100-100-100, SI-100-1k-100, SI-100-1k-225, GI-50-100-25 and GI-50-1k-25 as examples. **a-f**. Experimental prediction accuracies for 10% (4000) randomly selected test pairs for different datasets. These show the prediction accuracies at their location with regards to all 40,000 input-output pairs obtained. **g**. Examples of predicted patterns and their corresponding accuracies for different fibre lengths and acquisition methods. The drops in prediction accuracy at the beginning and the end of the measurement in **a-f** are caused by the reduced number of training data at the beginning and the end (detail discussions are in the Main Text).

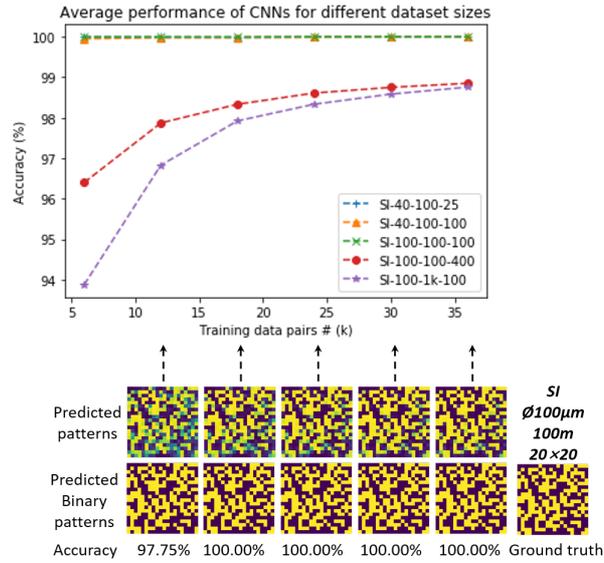

Fig. S5. The network performance at different sizes of randomly chosen training datasets (6000, 12000, 24000, 30000 and 36000). Insets below show the predicted patterns and their corresponding predicted binary patterns at different stages of the training, tested using Dataset SI-100-100-400 as an example.

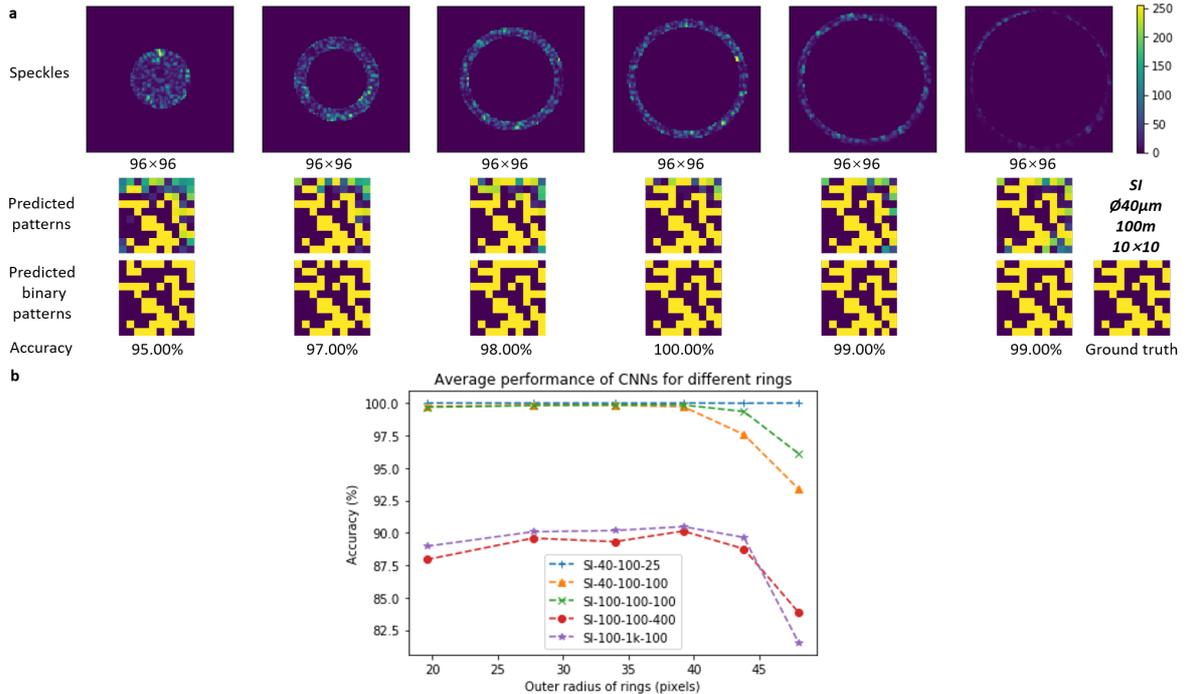

Fig. S6. **a**. Schematic for choosing the speckle area to be used as CNN's inputs (rings). The top row uses ring-shaped masks to preserve the same number of pixels used in each case. Therefore, each isolated area contains 1200 pixels approximately. Typically, lower-order modes are constrained to the centre of the fibre with higher-order modes towards the edges. Insets below show the predicted patterns and their corresponding predicted binary patterns in different cases of the training, tested using Dataset SI-40-100-100 as an example. A clear reconstruction of the complicated pattern has been observed even the dark blur like an edge is used as inputs. **b**. The average performance of the CNNs for different rings. Curves drawn here are based on four experiments over step-index MMFs.

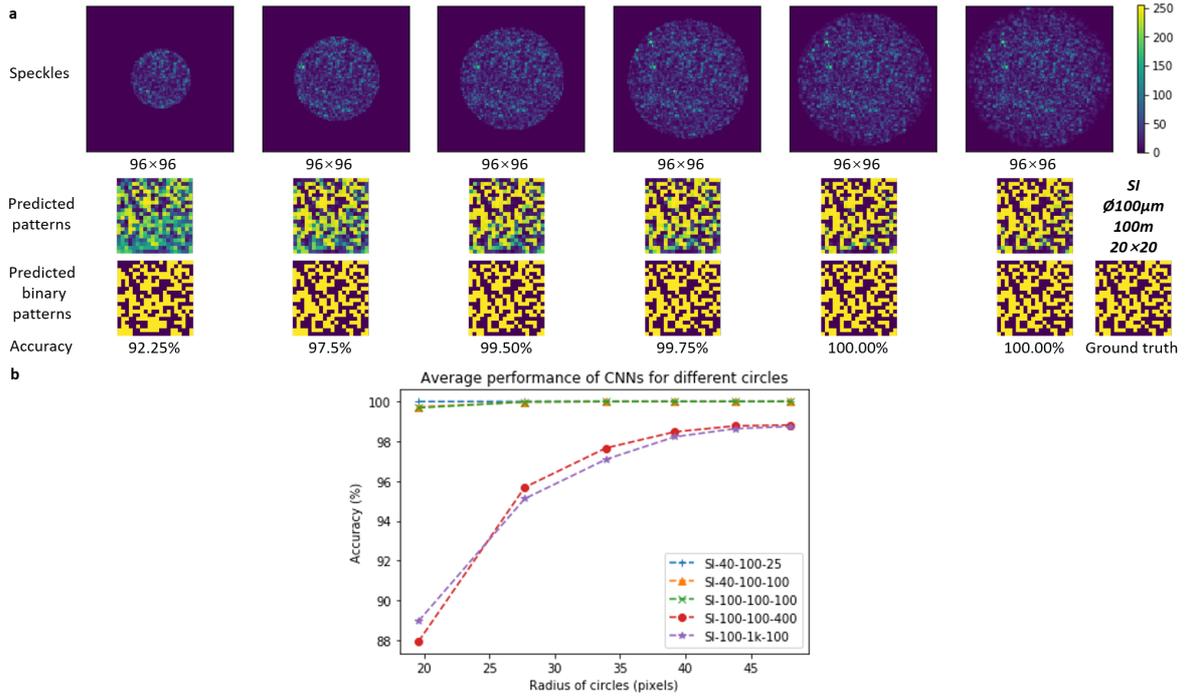

Fig. S7. **a**. Schematic for choosing the speckle area to be used as CNN's inputs (circles). The top row uses circular masks to isolate areas of the speckle pattern. The radiuses of the circles are equal to the outer radiuses of the rings shown in Fig. S6. Insets below show the predicted patterns and their corresponding predicted binary patterns in different cases of the training, tested using Dataset SI-100-100-400 as an example. **b**. The average performance of the CNNs for different circles. Curves drawn here are based on four experiments over step-index MMFs.

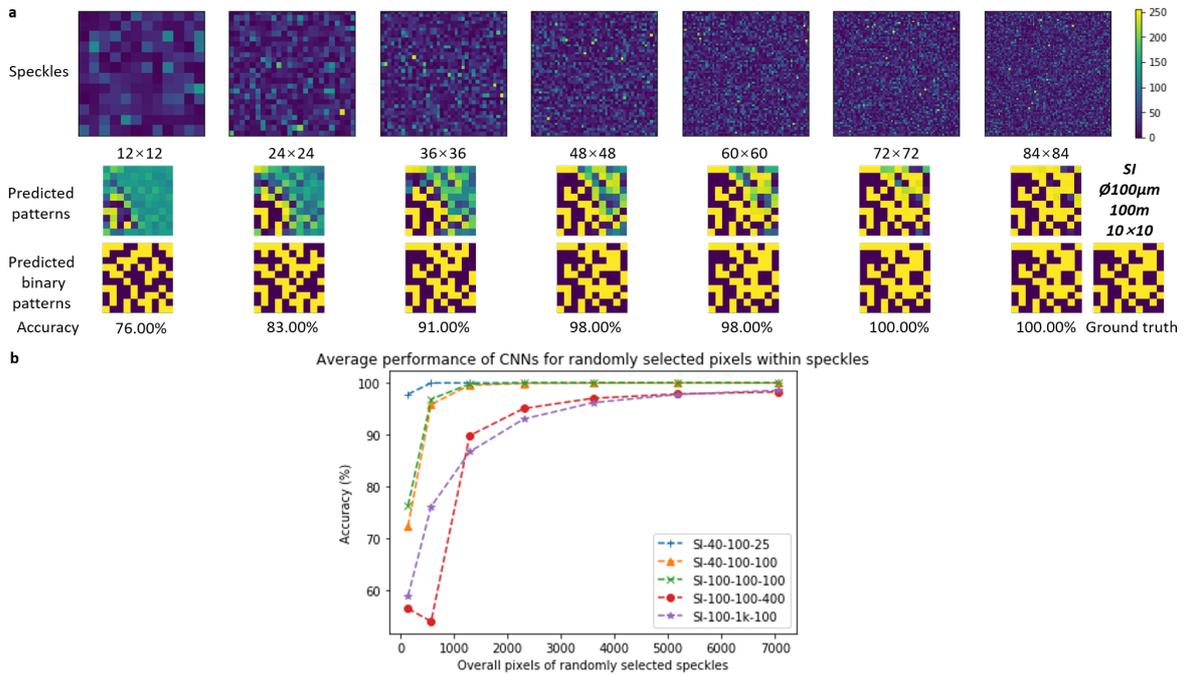

Fig. S8. **a**. Schematic for choosing the speckle area to be used as CNN's inputs (randomly selected patterns). The top row selects pixels randomly from the valid circular region to avoid the dark corners in each original speckle and shapes them in a square configuration. Insets below show the predicted patterns and their corresponding predicted binary patterns in different cases of the training, tested using Dataset SI-100-100-100 as an example. **b**. The average performance of the CNNs for different randomly selected speckle patterns. Curves drawn here are based on four experiments over step-index MMFs.

**Table S1. Average prequential accuracy, F1_score, MSE, PCC, SSIM (defined in Methods section in the main text) and updating runtime per step for Datasets: SI-40-100-25, SI-40-100-100, SI-100-100-100, GI-50-100-25, GI-50-100-100 and GI-50-1k-25. Each experiment was repeated 5 times and we report averaged results over these runs.**

| Datasets | Models | Accuracy (%) | F1_score | MSE | PCC | SSIM | Updating runtime |
|----------|--------|--------------|----------|-----|-----|------|------------------|

| | | | | | | | (*sec*) |
|---|---|---|---|---|---|---|---|
| SI-40-100-25 | Static | 99.77 | 0.9975 | 0.0027 | 0.9949 | 0.9931 | 502.9 |
| | SSL | 100.00 | 1.0000 | 0.0000 | 0.9999 | 0.9999 | 298.5 |
| SI-40-100-100 | Static | 91.61 | 0.9115 | 0.0622 | 0.8638 | 0.8582 | 410.7 |
| | SSL | 100.00 | 1.0000 | 0.0000 | 1.0000 | 1.0000 | 319.9 |
| SI-100-100-100 | Static | 99.65 | 0.9963 | 0.0035 | 0.9932 | 0.9931 | 401.1 |
| | SSL | 100.00 | 1.0000 | 0.0000 | 1.0000 | 1.0000 | 319.9 |
| GI-50-100-25 | Static | 66.43 | 0.5463 | 0.2749 | 0.3400 | 0.3087 | 467.4 |
| | SSL | 99.99 | 0.9998 | 0.0001 | 0.9998 | 0.9997 | 302.1 |
| GI-50-100-100 | Static | 87.07 | 0.8585 | 0.1049 | 0.7727 | 0.7877 | 384.2 |
| | SSL | 99.15 | 0.9913 | 0.0085 | 0.9831 | 0.9967 | 319.0 |
| GI-50-1k-25 | Static | 80.46 | 0.7744 | 0.1617 | 0.6106 | 0.6041 | 383.7 |
| | SSL | 99.99 | 0.9998 | 0.0001 | 0.9998 | 0.9997 | 295.9 |